\documentstyle[12pt,psfig]{article}

\newcommand{\AmS}{{\protect\the\textfont2
\renewcommand{\thesection}{\Roman{section}}
  A\kern-.1667em\lower.5ex\hbox{M}\kern-.125emS}}
 
\begin{document}
\rightline {DFTUZ 98/18}
\vskip 2. truecm
\centerline{\bf Baryonic thermal fluctuations in finite temperature QCD}
\vskip 2 truecm
\centerline { Vicente ~Azcoiti and Angelo ~Galante}
\vskip 1 truecm
\centerline {\it Departamento de F\'\i sica Te\'orica, Facultad 
de Ciencias, Universidad de Zaragoza,}
\centerline {\it 50009 Zaragoza (Spain).}
\vskip 3 truecm

\centerline {ABSTRACT}
\vskip 0.5truecm

\noindent
We show how, contrary to physical intuition, thermal fluctuations 
of physical states having a non-vanishing 
baryonic number can be fully neglected in the thermodynamics of $QCD$ at 
any physical temperature.
We also discuss on the consistency between our results and the $CPT$ theorem.
The last part of this article is devoted to 
comment some interesting physical features which follow from this result.

\vfill\eject
\baselineskip=24pt

The partition function of a quantum field theory at zero temperature is 
dominated by the contribution of the vacuum state, {\it i.e.}, 
the eigenstate of 
the quantum hamiltonian having minimum energy. By increasing the physical 
temperature, contributions to the free energy density of excited states 
become more and more relevant and in the infinite temperature limit the 
system fluctuates randomly and all the states have the same probability, 
no matter what energy they have.

In $QCD$, the gauge theory describing the strong interaction of particles, 
the $SU(3)$ gauge symmetry of the model implies that physical states are 
made up of a multiple of three number of quarks. Hence we have baryons and 
mesons in the spectrum of this model and a natural expectation is that 
thermal fluctuations of both kind of particles will be relevant at 
finite temperature. Notwithstanding that we will show in this letter how 
thermal fluctuations of physical states having a non-vanishing 
baryonic charge can be fully neglected in the thermodynamics of $QCD$ at 
any physical temperature \cite{VIC}. 
The last part of this article will be devoted to 
discuss some interesting physical features which follow from this result.

\vskip 0.3truecm
\noindent
{\bf 1. Analytical results}
\vskip 0.3truecm

Due to a global U(1) symmetry of $QCD$, the vector current 
$\bar\psi\gamma_{\nu}\psi$ is a conserved current. The integrated 
fourth component of this current 
 
\begin{equation}
N = \int d^{3}x \bar\psi(x)\gamma_{0}\psi(x)
\label{1}
\end{equation}

\noindent
defines a dynamical invariant, the baryonic charge or quark number operator. 
After quantization the quark number operator commutes with the hamiltonian and 
physical states can therefore be characterized by its energy and 
baryonic charge. Hence 
the Hilbert space of physical states can be decomposed as a direct sum 
of Hilbert subspaces, each one of them corresponding to a given eigenvalue 
of the quark number operator.

Taking into account this decomposition of the Hilbert space of physical 
states we can write the partition function of $QCD$ at finite temperature

\begin{equation}
Z=Tr\left( e^{-\frac{H}{T}}\right)
\label{2}
\end{equation}

\noindent
as a sum of canonical partition functions

\begin{equation}
Z=\sum_{k}Tr_{k}\left( e^{-\frac{H}{T}}\right) ,
\label{3}
\end{equation}

\noindent
each one of the $Tr_{k}\left( e^{-\frac{H}{T}}\right)$ being the partition 
function 
computed in the baryonic sector of the Hilbert space corresponding to a given 
eigenvalue $k$ of the quark number operator.

In order to establish the connection between this hamiltonian approach and the 
path integral formulation in the lattice regularization scheme, and to identify 
in the latter formalism what are the different baryonic sector contributions 
to the partition function, the simplest way is to introduce a chemical 
potential $\mu$ since it couples directly with the quark number operator $N$.
The Grand Canonical Partition Function $(GCPF)$ is given then by the 
following expression

\begin{equation}
Z_{GC} = Tr\left( e^{-{1\over{T}}(H-\mu N)}\right) .
\label{4}
\end{equation}

Using now the previous decomposition of the Hilbert space in different 
baryonic sectors we can write the $GCPF$ as 

\begin{equation}
Z_{GC} = \sum_{k}e^{{\mu\over{T}}k}Tr_{k}\left( e^{-\frac{H}{T}}\right) , 
\label{5}
\end{equation}

\noindent
which is the standard decomposition of the $GCPF$ as a sum of canonical 
partition functions. Setting the chemical potential $\mu=0$ we recover 
expression (3).

The standard way to introduce a chemical potential in lattice regularized 
$QCD$ is to multiply all temporal link variables pointing forward 
(backward) by $e^{\mu}$ ($e^{-\mu}$) \cite{KOGUT}, \cite{HASENFRATZ}. 
The previous decomposition of the $GCPF$ as a sum of canonical partition 
functions corresponds on the lattice to the Polyakov loop expansion of 
the partition function 

\begin{equation}
Z_{GC} = \bar{a}_{3V_x} e^{3V_{x}{\mu\over{T}}} 
+\bar{a}_{3(V_x-1)} e^{3(V_{x}-1){\mu\over{T}}}+.....
+\bar{a}_{0}+.......+\bar{a}_{-3V_x} e^{-3V_{x}{\mu\over{T}}},
\label{6}
\end{equation}

\noindent
where $V_x$ is the lattice spatial volume, the temperature $T$ is the inverse 
lattice temporal extent and the indices in the sum are numbers of vanishing 
triality as a consequence of the fact that physical states are made up 
of a multiple of three number of quarks. This last property can also be 
derived from the invariance of the pure gauge action under $Z(3)$ global 
transformations. 

The relation between the integrated 
coefficients $\bar{a}_{k}$ in (6) and the contribution $a_{k}(U)$ 
from a single gauge configuration is 

\begin{equation}
\bar{a}_{k}= \int [DU] a_{k}(U) e^{-\beta S_{G}(U)}.
\label{7}
\end{equation}

\noindent
where $S_{G}(U)$ is the standard pure gauge Wilson action, $\beta$ the 
square of the inverse gauge coupling and $a_{k}(U)$ is the contribution
to the determinant of the Dirac-Kogut-Susskind operator containing a net
number of Polyakov loops equal to $k$. 

Expression (6) for the partition function has been written assuming the 
Kogut-Susskind regularization for the fermionic degrees of freedom. Under 
such a regularization, $3V_x$ is the maximum number of Polyakov loops 
which can appear in the fermion determinant for a lattice with $V_x$ 
spatial points. Equivalently $3V_x$ is also the maximum number of 
three-colored quarks of the Kogut-Susskind type we can put in a finite 
discrete space with $V_x$ points following the Fermi-Dirac statistics.

From the comparison of expressions (5) and (6) we can conclude that the 
coefficient $\bar{a}_{k}$ is equal to the canonical partition 
function $Tr_{k}(e^{-\frac{H}{T}})$ computed in the baryonic sector of 
the Hilbert space having a baryonic number equal to $k\over{3}$ 
\cite{KOGUT}. 

In order to analyze the relevance of the different coefficients $\bar{a}_{k}$ 
for different values of the baryon number $k\over{3}$, we will consider 
the fermion determinant for a given gauge configuration in  
presence of an imaginary chemical potential 
$\mu = {{i\eta}\over{L_t}}$ \cite{VIC}, 

\begin{equation}
\det \Delta(\eta, U)= \sum_{k} a_{k}(U) e^{ik\eta}, 
\label{8}
\end{equation}

\noindent
where the sum in (8) is over all integer numbers between $-3V_x$ and $3V_x$. 

The antihermiticity and $\gamma_5$ anticommuting 
properties of the massless Dirac operator are preserved after the inclusion 
of an imaginary chemical potential. 
It follows that the positivity of the fermion determinant 
is also preserved and

\begin{equation}
\det \Delta(\eta, U)\geq{0} 
\label{9}
\end{equation}
 
\noindent
for any gauge configuration and any $\eta$. 
This property of positivity will be fundamental in order 
to derive bounds for the partition functions $\bar{a}_{k}$ which will 
allow us to get interesting conclusions on the relevance of the different 
baryonic sectors.
In a similar way Vafa and Witten used the positivity of 
the effective fermionic 
action in vector-like theories to get bounds 
which drive to the impossibility to break spontaneously vector-like 
global symmetries as isospin or baryon number conservation \cite{WITTEN}.

A simple Fourier transformation in (8) gives 

\begin{equation}
a_{k}(U) = {1\over{2\pi}}\int e^{-ik\eta} \det \Delta(\eta, U) d\eta 
\label{10}
\end{equation}

\noindent
and, making use of the positivity relation (9), the following inequalities 
can be derived

\begin{equation}
a_{0}(U)\geq{0}, \qquad
|a_{k}(U)| \leq{a_{0}(U)} \qquad \forall k. 
\label{11}
\end{equation}

These relations, which hold for any gauge configuration,  
imply for the integrated coefficients or canonical partition 
functions $\bar{a}_{k}$ the inequalities $\bar{a}_{k} \leq{\bar{a}_{0}}$, 
{\it i.e.}, 

\begin{equation}
{{Tr_{k}\left( e^{-\frac{H}{T}}\right) }
\over
{Tr_{0}\left( e^{-\frac{H}{T}}\right) }}\leq{1}. 
\label{13}
\end{equation}
 
The relation (\ref{13}) 
between canonical partition functions induces interesting 
physical consequences in the thermodynamics of this model. To analyze them 
let us consider the free energy density,

\begin{equation}
f = {T\over{V_x}} \log Z. 
\label{14}
\end{equation}

Using the obvious relation $\bar{a}_k = \bar{a}_{-k}$ the partition 
function at zero chemical potential can be written as

\begin{equation}
Z = \bar{a}_0 \left( 1+ {{2\bar{a}_3}\over{\bar{a}_0}}+
...........+ {{2\bar{a}_{3V_x}}\over{\bar{a}_0}}\right) , 
\label{15}
\end{equation}

\noindent
a relation which gives for the free energy density the following expression

\begin{equation}
f = {T\over{V_x}} \log \bar{a}_0 + 
{T\over{V_x}} \log \left( 1+ {{2\bar{a}_3}\over{\bar{a}_0}}+
...........+ {{2\bar{a}_{3V_x}}\over{\bar{a}_0}}\right).
\label{16}
\end{equation}

The first contribution in (\ref{16}) is the free energy 
density of the vanishing 
baryon number sector. The second term contains the contribution of all the 
other non-vanishing baryon number sectors. Relation (12) implies that the 
only contribution surviving the thermodynamical limit is the first one. The 
free energy density of $QCD$ at any physical temperature $T$ is given 
therefore by

\begin{equation}
f = {T\over{V_x}} \log \bar{a}_0 = 
{T\over{V_x}} \log \left( Tr_{0}\left( e^{-\frac{H}{T}}\right)\right) . 
\label{17}
\end{equation}
 
The previous expression implies that all baryonic thermal fluctuations can 
be neglected in the thermodynamics of $QCD$ (at zero chemical
potential). Notwithstanding that, this result 
does not necessarily imply that the probability $p_{k}$ to get a sector 
with non-vanishing baryon number vanishes. What happens is that if 
${{p_{k}}\over{p_{0}}} = {{\bar{a}_{k}}\over{\bar{a}_{0}}}$ is a 
non-vanishing number in the infinite volume limit, 
sector $k$ has the same free energy density as sector $0$ 
and therefore the same thermodynamical properties. 

The CPT theorem \cite{PCT} states that the vacuum state is CPT invariant.
Since the baryon number density is an order parameter for the CPT
symmetry, the theorem enforces the probability distribution function
of the baryon number density to be a $\delta$ function centered at the
origin. Our results are therefore consistent with the CPT theorem but
also give some further information: the partition function of QCD
at any physical temperature can be computed not only in the vanishing 
baryonic density sector but also in the vanishing baryon number sector
{\it i.e.}, all baryonic thermal fluctuations can be neglected.

This is something different from the standard statistical mechanics
result that relative fluctuations go to zero as the thermodynamical
limit is approached. In fact we state that all properties related to 
the free energy density can be 
derived considering the zero baryon number sector only. Clearly the
result does not apply to the derivatives respect to the chemical
potential {\it i.e.}, the baryonic susceptibility at zero chemical potential
can not be evaluated from the vanishing baryon number sector only.

\vskip 0.3truecm
\noindent
{\bf 2. Some interesting physical implications}
\vskip 0.3truecm

Once we have demonstrated that thermal fluctuations of physical states with 
non-vanishing baryon number can be neglected in the thermodynamics of $QCD$, 
we will devote the last part of this article to the discussion of some 
interesting physical implications of this result.

{\bf i)} The first interesting feature is the possibility of analyzing the 
deconfining phase transition at finite temperature by means of an order 
parameter. The mean value of the Polyakov loop or Wilson line \cite{POLY}
is related 
to the energy of an isolated quark in the fluctuating background at a 
given physical temperature $T$. The standard wisdom is that the mean value of 
this operator is not an order parameter in the full theory with dynamical 
fermions since the determinant of the Dirac operator, which appears in the 
integration measure, breaks explicitly the Polyakov symmetry. The possibility 
of using the mean value of the Polyakov loop as an exact order parameter 
in the full theory with dynamical fermions, which was pointed out in early 
works \cite{VIEN1}, \cite{VIEN2}, becomes apparent 
in the formalism here developed. In 
fact $a_{0}(U)$, the contribution to the fermion determinant containing 
a vanishing net number of Polyakov loops, is invariant under Polyakov 
transformations. Even more, $a_{0}(U)$ is also independent of the 
boundary conditions for the fermion field in the temporal direction. 
Periodic or antiperiodic boundary conditions give rise to the same 
thermodynamical properties. 

{\bf ii)} Theoretical prejudices based on simple 
models tell us that the high density 
phase of $QCD$ should be similar to the high temperature phase in the sense 
that both phases would be described by a free quark-gluon plasma. However 
the physical background in the two limiting cases, high temperature at 
vanishing chemical potential and high baryonic density at vanishing 
temperature, shows important differences. In the first case the ground state 
is dominated by thermal fluctuations of physical states with vanishing 
baryon number whereas in the second case, physical states with a high 
baryonic density are the only relevant contributions. Therefore even if 
the theoretical scenario of a quark-gluon plasma phase describes actually 
the two phases, we should expect important phenomenological differences 
between the two limiting extremes of the $(\mu, T)$ phase diagram.

{\bf iii)} The results here obtained allow also to solve 
the inconsistencies found 
in quenched $QCD$ simulations at finite temperature. As it was shown a 
few time ago \cite{UNO}, \cite{MISHA} quenched simulations of $QCD$ at finite 
temperature and in the broken deconfined phase gave inconsistent results 
for the chiral condensate and different values for the critical temperature 
of the chiral transition were found, depending on the phase of the Polyakov 
loop in which the dynamics settles. These results, which put in question the 
validity of the quenched approximation, are not unexpected. In fact the 
integration measure is invariant under Polyakov transformations in this 
approximation but the chiral condensate operator is not. A value of the 
chiral condensate depending on the $Z(3)$ vacuum in which the dynamics settles 
 in the broken phase is a rather expected result on physical grounds. 
However, as previously shown, the free energy density of $QCD$ can be computed 
neglecting all the contributions to the partition function of non-vanishing 
baryon number sectors. We can take $a_{0}(U)$ instead of $\det\Delta(U)$ in 
the integration measure of full $QCD$ and consistently the logarithmic mass 
derivative of $a_{0}(U)$ instead of the trace of the inverse Dirac operator 
as chiral condensate operator. Within this prescription, the chiral order 
parameter is invariant under Polyakov transformations and its mean value is 
therefore the same in each of the three possible 
$Z(3)$ vacuums of the quenched broken phase.

{\bf iv)} Another topic in which the results here reported 
could give rise to an 
interesting improvement is the one concerning finite size effects and 
Monte-Carlo convergence in numerical simulations of finite temperature $QCD$. 
As previously discussed, most of the contributions of the different 
baryon number sectors, specially those corresponding to a non-vanishing 
baryon number density, will have a vanishing probability in the 
thermodynamical limit. They are pure finite size effects and a natural 
expectation is that with our prescription of taking $a_{0}(U)$ instead of 
$\det\Delta(U)$ in the integration measure, we will approach the 
thermodynamical limit faster. But even if some of these contributions have 
a non-vanishing probability, our prescription of taking $a_{0}(U)$ in the 
integration measure should improve the Monte-Carlo convergence in numerical 
simulations of finite temperature $QCD$. In fact the coefficient $a_{0}(U)$ 
has more symmetries than the determinant of the full Dirac operator and, as 
we have learned from the $GCPF$ computation in the investigations of $QCD$ 
at finite density \cite{BARBOUR}, \cite{NOI} $a_{0}(U)$ is less 
fluctuating than $\det\Delta(U)$. 
This last feature is the manifestation of what is known as 
the sign problem in finite density $QCD$ \cite{BIELEFELD}, a problem 
that has delayed enormously the progress in this field.

The central coefficient $a_{0}(U)$ is positive definite for any gauge 
configuration but unfortunately this is not the case for almost all the 
other coefficients. Most of them fluctuate violently with a rather constant 
absolute value but with an almost randomly distributed phase. Things here 
are very similar to what happens in the unpleasant situation we find when 
try to compute a small number by averaging numbers which are several orders 
of magnitude larger than their mean value. We would need a prohibitive 
statistics, increasing exponentially with the lattice volume in the GCPF 
computations \cite{NOI}, \cite{BIELEFELD}, 
to get significant numbers. The contribution of most of the 
non-vanishing baryon number sectors to the partition function of finite 
temperature $QCD$ at actual statistics behaves therefore like a 
noise, and a natural expectation is that convergence of any Monte-Carlo 
procedure should improve by killing this noise.

The exact computation of $a_{0}(U)$ is very hard. The best way at present 
is through the construction of the quark propagator matrix \cite{GIBBS} 
and requires the exact diagonalization of a non hermitian matrix 
for each gauge configuration. 
However, as follows from the analysis here developed, there 
is a simpler way to implement things in practical standard simulations of 
$QCD$ at finite temperature. It consists in the introduction of an extra 
abelian degree of freedom $e^{i\eta}$ coupled to all temporal links and 
having a dynamical character. The integration over this new dynamical degree 
of freedom will kill all the undesidered contributions to the partition 
function.

\newpage
\noindent
{\bf 3. Wilson fermions}
\vskip 0.3truecm

We have shown in the previous sections of this paper how the thermodynamics 
of QCD, when regularized in a space-time lattice and using staggered fermions, 
is controlled by the contribution to the fermion determinant with no net 
number of Polyakov loops, {\it i.e.}, by the thermal fluctuations of 
physical states with vanishing baryon number.

The two main ingredients to get this result are the conservation of 
baryonic charge in QCD and the inequalities (12) which tell us 
that the partition function at fixed baryon number reaches its maximum value 
in the Hilbert subspace corresponding to zero baryon number. The first one 
of the two ingredients is independent of the lattice regularization for the 
Dirac operator. The second one however is based on the positivity of the 
determinant of the Dirac operator for any gauge configuration and any value 
of the extra-abelian degree of freedom $e^{i\eta}$.

Since we have made use of the hermiticity and chiral properties of the 
Dirac-Kogut-Susskind operator, it is natural 
to ask whether our result applies to any fermion regularization or rather 
it is related to the presence-absence of the chiral anomaly.

Let us say from the beginning that even if we have not yet a definite 
answer to this question, there are strong indications suggesting that the 
chiral anomaly does not play any relevant role here. These indications come 
from the analysis of the properties of the fermion determinant for 
Wilson fermions. As well known, the Dirac-Wilson operator $\Delta$ can 
be written as 

\begin{equation}
\Delta = I - \kappa M, 
\label{18}
\end{equation}
where $\kappa$ is the hopping parameter and the matrix M verifies the 
following chiral relation

\begin{equation}
\gamma_{5} M \gamma_{5} = M^{+}. 
\label{19}
\end{equation}

Equation (\ref{19}) implies that if $\lambda$ is eigenvalue of M, 
$\lambda^{*}$ is 
also eigenvalue of M, {\it i.e.}, 
the fermion determinant is always real. However 
it could be negative and in fact this unpleasant situation has been found 
for some gauge configurations in numerical simulations of the Schwinger model 
done in the unphysical strong coupling region \cite{NOS}. However the unitary 
character of the gauge group implies that all the eigenvalues are upper 
bounded by the relation

\begin{equation}
|\lambda | \leq{8} 
\label{20}
\end{equation}
which implies that for $\kappa \leq{1/8}$, $\det \Delta \geq{0}$. It is easy 
to verify that under the previous condition $\kappa \leq{1/8}$, the 
positivity of $\det \Delta$ also holds in the presence of the extra-abelian 
degree of freedom previously introduced.

In other words, all the results of this paper can be extended in a 
straightforward way to Wilson fermions if we impose the restriction 
$\kappa \leq{1/8}$, {\it i.e.}, the hopping parameter region associated to a 
positive bare fermion mass.

\newpage
\noindent
{\bf Acknowledgements}
\vskip 0.3truecm

This work has been partially supported by CICYT (Proyecto AEN97-1680). 
The authors thank also INFN for financial support.


\newpage
\vskip 1 truecm
\begin{thebibliography}{9}

\bibitem{VIC}
V. Azcoiti, 
Theory of Elementary Particles ( Proceedings of the 31st International 
Symposium Ahrenshoop) WILEY-VCH \rm (1997) 284. 

\bibitem{KOGUT}
J.B. Kogut, H. Matsuoka, M. Stone, H.W. Wyld, S. Shenker, J. Shigemitsu, 
D.K. Sinclair,
{\it Nucl. Phys.} {\bf B225} [FS9] \rm (1983) 93. 

\bibitem{HASENFRATZ}
P. Hasenfratz, F. Karsch, 
{\it Phys. Lett.} {\bf 125B} \rm (1983) 308. 

\bibitem{WITTEN}
C. Vafa, E. Witten, 
{\it Nucl. Phys.} {\bf B234} \rm (1984) 173. 

\bibitem{PCT}
F. Streater, A.S. Wightman, PCT, Spin and Statistics and All That, Benjamin, 
New York (1964).

\bibitem{POLY}
A.M. Polyakov, {\it Phys. Lett.} {\bf B72} (1977) 477.

\bibitem{VIEN1}
M.~Oleszczuk and J.~Polonyi, 
{\it Ann. of Phys.} {\bf 227} \rm (1993) 76.

\bibitem{VIEN2}
M. Faber, O.A. Borisenko, S. Mashkevich, G.M. Zinovjev. 
{\it Nucl. Phys.} {\bf B} {\it (Proc. Suppl.)} {\bf 42} (1995) 484; 
M. Faber, O.A. Borisenko, G.M. Zinovjev. 
{\it Nucl. Phys.} {\bf B444} \rm (1995) 563.

\bibitem{UNO}
S. Chandrasekharan, N. Christ, {\it Nucl. Phys.} {\bf B} 
{\it (Proc. Suppl.)} {\bf 47} \rm (1996) 527.

\bibitem{MISHA}
M.A. Stephanov, {\it Phys. Lett.} {\bf B375} (1996) 249.

\bibitem{BARBOUR}
I.M. Barbour, S.E. Morrison, E.G. Klepfish, J.B. Kogut, M.P. Lombardo,
{\it Phys. Rev.} {\bf D56} \rm (1997) 7063.

\bibitem{NOI}
R. Aloisio, V. Azcoiti, G. Di Carlo, A. Galante, A.F. Grillo,
{\it Phys. Lett.} {\bf B428} (1998) 166; 
{\it Phys. Lett.} {\bf B435} (1998) 175.

\bibitem{BIELEFELD}
I.M. Barbour, to appear in the proceedings of the 
Conference "QCD at finite baryon density",
Bielefeld, April 1998.

\bibitem{GIBBS}
P.E. Gibbs,
{\it Phys. Lett.} {\bf B172} \rm (1986) 53. 

\bibitem{NOS}
V. Azcoiti, G. Di Carlo, A. Galante, A.F. Grillo, V. Laliena, 
C.E. Piedrafita, {\it Phys. Rev.} {\bf D53} (1996) 5069.


\end{thebibliography}
\end{document}